# DRMS Co-design by F4MS

**Aissam BERRAHOU, Mourad RAFI and Mohsine ELEULDJ**

**Computer Department, Mohammadia School of Engineers**
**Rabat, BP 765, Avenue Ibn Sina Agdal, Morocco**

**Abstract**
In this paper, we present Digital Rights Management systems (DRMS) which are becoming more and more complex due to technology revolution in relation with telecommunication networks, multimedia applications and the reading equipments (Mobile Phone, IPhone, PDA, DVD Player,..). The complexity of the DRMS, involves the use of new tools and methodologies that support software components and hardware components coupled design. The traditional systems design approach has been somewhat hardware first in that the software components are designed after the hardware has been designed and prototyped. This leaves little flexibility in evaluating different design options and hardware-software mappings. The key of co-design is to avoid isolation between hardware and software designs to proceed in parallel, with feedback and interaction between the two as the design progresses, in order to achieve high quality designs with a reduced design time. In this paper, we present the F4MS (Framework for Mixed Systems) which is a unified framework for software and hardware design environment, simulation and aided execution of mixed systems. To illustrate this work we propose an implementation of DRMS business model based on F4MS framework.

## 1. Introduction

In the heart of the digital economy, the Digital Rights Management (DRM)[1], [2], [3], [4] must fulfill the requirements of access control, use and diffusion of any digital contents from computer, Mobile Phone or other equipment through internet or telecommunication network.

Systems that provide digital rights management (DRM) [5] are very complex, extensive and not flexible: DRM technologies must support a diversity of devices (Mobile, PDA, PC, ..), users, platforms (Media player, web server,..), and media (audio, video, image, text, application, cloud computing..), and a wide variety of system requirements concerning security, flexibility, manageability, reuse, maintainability, interoperability.

Existing tools design [6], [7] for DRM systems are limited only to software design in isolation with the hardware design which is an important step in such systems. A reading equipment with less performance cannot deal with voluminous media content even if we have a high media player performance.

The DRM community [8], [9], [10], [11], [12] needs a co-design framework to design and develop a high performance, flexible, reuse, maintainable DRM systems with less cost time to deal with rapid growing DRM market.

The main focus for the development of mixed systems, using co-design framework, remains in the partitioning of tasks. However the major challenges of mixed systems are their development and use, taking into account the coexistence between software and hardware, as well as the multiple and complex interactions between various components.

The main goal of this work is not only the proposition of a design methodology (flexible) for the specification and the partitioning of software / hardware, but also provides a framework for the implementation of systems incorporating both hardware and software components, as well as the proposition of a general model for design and execution of mixed systems.

This paper is organized as follows. Firstly, section 2 presents the standard DRMS Architecture. Then Section 3 presents the F4MS framework for the design and execution of mixed systems. It describes the design methodology and the general model of these systems. Then, Section 4 presents the design and implementation of a mixed system for DRMS business Model. Finally, conclusion is presented in section 5.

## 2. Standard DRMS architecture

A standard DRMS architecture (figure1) is composed by three components: Creation, distribution and consumption of the digital content [6]:

**Consumption:** Consumers want to be able to browse the content catalog of the on-line DRM system where the content at stake can be obtained. Since consumers also need a license, they must be able to select a license type and view the usage rules associated with it. Generally, consumers first have to pay, one way or another; different business models should be possible (e.g. subscription, pay-per-license, or pay-per-use). When time-based licenses expire, it must be possible to update them, which may also require some financial transaction. Consumers also want to browse their obtained licenses locally and view the usage rules in a human readable format. Finally, consumers want to consume the protected content, according to the usage rules associated with the corresponding license. In order to fetch licenses (and sometimes also protected content), consumers need to authenticate to the on-line DRM system.

**Creation:** Content Producers want to easily compose a contract. Both content and contract must be submitted to the on-line DRM system. After some time, they may want to update the contract or maybe even cancel it, i.e. stop the distribution of the content. Content producers expect a financial compensation from the DRM service for the trade of their content. Therefore, they want to receive statistical information from the DRM service about the number of downloads or content usage patterns. In order to query or submit content to the on-line DRM system, content producers need to authenticate themselves.

**Distribution and publishers:** When one or more DRM clients are no longer secure, their right to consume content must be revoked. It may also be necessary to update some parts of the DRM system (and the DRM client). Content publishers may want an overview of system usage patterns. When content is found mass-distributed, the source of abuse must be identifiable.

An example of standard DRMS is as follow (figure1):

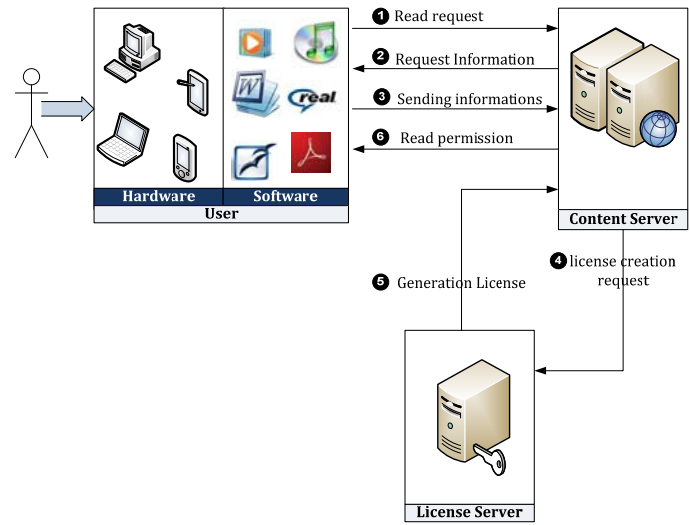

Fig. 1 Standard DRMS architecture

This standard DRMS work as follow:

❶ User requests a digital content.
❷ The content server demand to user to fill some information.
❸ User sends the information.
❹ The Content server requests license generation to license server.
❺ The license server generate license and send it to the content server.
❻ The content server gives to user authorization to read the digital content.

The correspondent UML sequence diagram is illustrated by the figure 2.

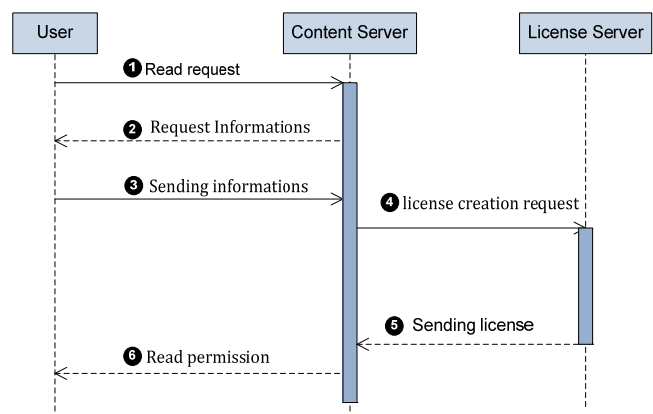

Fig. 2 Sequence diagram

After analysis of DRMS, we find that they consist of two parts in constant interaction: hardware components (PC, iPhone, PDA, License server, ...) and software components (media player, web server, license generator application...).

Existing tools design for DRM systems [1], [2] are limited only to software design in isolation with the hardware design which is an important step in such systems. A reading equipment with less performance cannot deal with voluminous media content even if we have a high media player performance.

The DRM community needs a co-design framework to design and develop a high performance, flexible, reuse, maintainable DRM systems with less cost time to deal with rapid growing DRM market.

In the next section, we will present a co-design framework which is a unified framework for software and hardware design environment, simulation and aided execution of mixed systems. The co-design framework used is called F4MS (Framework For Mixed Systems)

## 2. Framework for mixed systems

F4MS (Framework for mixed systems) [13] represents an extension of the TI4CS framework [14] (Tools Integration for Complex Software) which not include only software components but also hardware components. This new extension is dedicated to the design of systems based on execution graph. The design of this type of systems by the F4MS framework is made in two stages: the first stage consists in defining all the components (Software and\or hardware) necessities for the system to realize. The second stage consists in integrating these components in the form of a execution graph, which contains the description of the final application, i.e how the components are organized (position of the component in the graph), and how they will interoperate with each other.

In the following, we present the characteristics of F4MS framework and the design methodology.

### 2.1 Features of F4MS

F4MS has the same characteristics of TI4CS framework [13], [14], [15] at the same time gives the possibility of integrating two types of components completely different in terms of design and architecture.

We quote here a number of characteristics that we have considered during the design stage of the new version of that framework:

**A modeling of several levels of abstraction:** The specification can use levels of abstraction combined to accelerate simulation, because in certain cases, any detail is not necessary to test the concept of a design and some of its functionalities.

**A separation between the communication model and the treatment model describing the system:** the size (format) has to allow a refinement of the communication network between independent modules and the optimization of the internal behavior of system components.

**Heterogeneity** is the possibility to use multiple programming languages and hardware architectures (FPGA, ASIC, etc...), each one for a different part of the system. However, and since each one of these elements has these properties which can be exploited for the optimized analysis and implementation. These properties are different from a one model to another which inhibits the analysis and the optimization of the whole system beyond the limits of the language and hardware architecture, and thus the platform must resolve this problem.

**Distributed Validation:** it makes it possible to distribute system modules to be validated through a network. It is then possible to simulate a whole system while using a suitable simulator for each development team. The simulators do not any more need to be all grouped on a machine or a particular site which generally requires a more computing power. The teams can then collaborate in the level of simulation and of the development. Of course, the simulation performances are dependent on the network used and its load during the simulation of the system.

### 2.2 Design methodology

The design methodology of mixed architectures [16], [18], grouped the technical aspects and the organization aspects. It coordinates the use of several combined tools of conception and the cooperation of several aspects bound at all level of a system development. The teams of the software and the teams of the hardware can work in parallel, in an environment of cooperation and collaboration which reduces the development cycle of the conception [18], [19], [20].

The designers must take several decisions to clarify the details of this architecture [21], [22], in a consistency that allows the best compromise driving performances / area / consumption / flexibility / reusability / manageability/ quality/ interoperability/cost design/time to market.

The main design steps are summarized below:

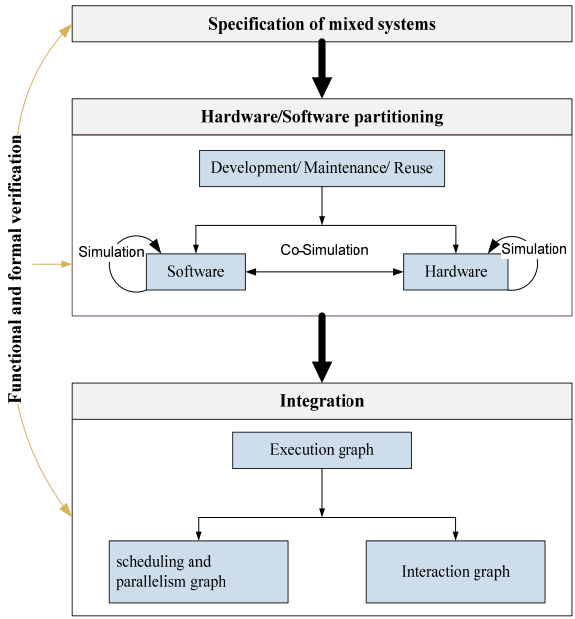

### 2.2.1 System specification

They are high level specifications of abstraction and allow expressing the needs of a designed system:

**Functional specifications** describe with exactitude what the system will have to carry out (details of the operations to be done, relations between inputs and outputs, and results to produce).

**Non-functional Specifications** describe the conditions under which the system must operate: meet performances, consumption, average yield, the cost of manufacture, etc.

### 2.2.2 Architecture

This stage consists of three steps [14]:

**The partitioning or splitting of the system to software/hardware component based on the estimation performance:** explore the alternatives of design to identify those which adapt best to the constraints of the system. This step carries out the transposition of the functions of the system on software and hardware components. The components are software entities, programmable processors, FPGAs, and memories.

**Development, Maintenance, Reuse:** This step consist of implementing new components, maintain existing components and trying to adapt them to the context, or / and reuse of components developed by other agencies.

**The co-simulation** is an important step in validating the behavior of a component after the Software/Hardware partitioning.

### 2.2.3 Integration

The execution graph [13] of F4MS is a workflow for the description of mixed systems architectures, based on software / hardware components. He allows describing a system as being a set of components (monolithic or composite) which implement interfaces, connectors (interconnections between components) and their compositions.

A component of execution graph presented as a calculation unit, or a data warehouse. An interface specifies the services which the component provides. The connectors model the sequencing and interaction between components through their interfaces. A composition represents a graph of components, connected between them using connectors.

The execution graph (Fig 3) is consists of two graph: the scheduling and parallelism graph (to organize the components and describe the sequence of execution) and the interaction graph (to ensure interoperability between components), it also includes information about the parameter setting and the configuration of the properties of execution with each component.

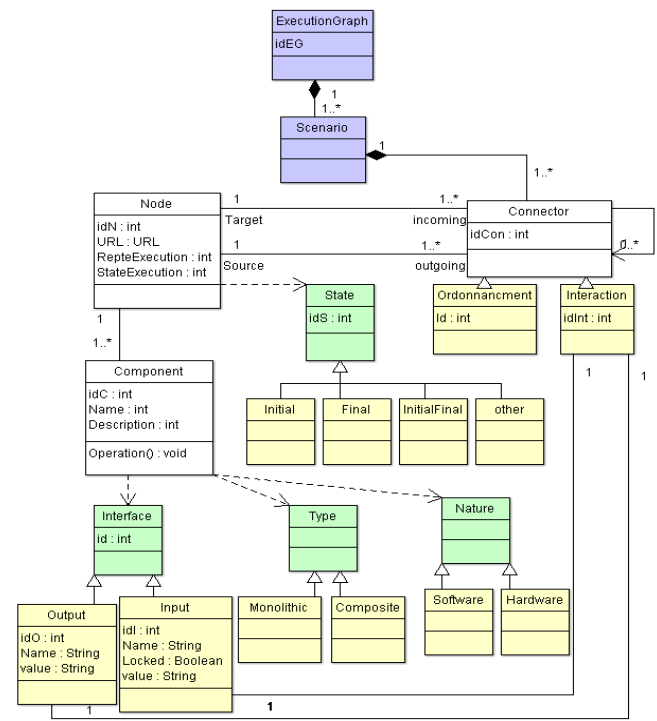

Fig. 3 Meta-model of mixed systems

## 3. Application

In this section we will describe the design phases of DRMS business model using F4MS co-design framework.

The DRMS business model proposed is illustrated by the UML sequence diagram (Fig 4):

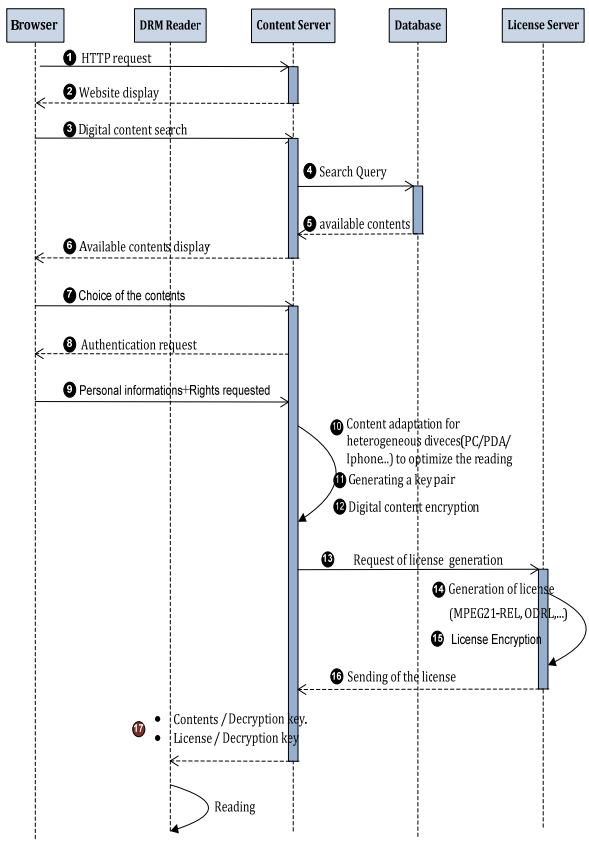

Fig. 4 DRMS business model

### 3.1 DRMS specification

The design that we proposed consists of two parts:

**The non functional part** of the application which provides the security constraints[23] to be respected (needs analysis in terms of cryptographic techniques(AES, DES, TDES, RSA,…), and in technological choices (REL: ORDL[24]/MPEG21-REL[25], [26], [27], OpenSSL, IPSec, VPN) for the implementation of the DRMS), the objectives and the advantages of these systems compared to what already exists on the market in term of quality, speed of development, modularity, maintainability, portability, feasibility and most importantly ease of use.

**The functional part** which is to define the organizational structure (the scheduling and parallelism graph) of DRM systems in scenarios which represents the same application model and describes the order execution of components. We note that this phase of specification is made after the second stage of the design methodology that we will present later.

### 3.2 Architecture

#### 3.2.1 DRMS Partitioning

Partitioning is an operation charged by different skill levels (integrators) and having an expertise in specification, optimization, and integration. Generally the objective aimed by this phase of conception is to obtain a homogeneous architecture for low cost and which satisfy the constraints of execution of the application.

When partitioning, it is to answer the following questions:
– What are the components necessary to establish the DRM solutions?
– With which technology (software or hardware) must be used to implement each component?
– Which is the profit (performance, surface, and cost) obtained after an operation of refinement?
– When we can replace software component by hardware component and vice versa to handle performance requirement?

To answer these questions, we have studied the needs, which must answer a certain criterion that was listed.

#### 3.2.2 Components development

The study of the needs led us to identify three types of components that we saw necessary to the establishment of DRMS:
– User components.
– Content server components.
– License server components.

Table 1: Components list

| Components | | Description |
|---|---|---|
| User | 1 | browser |
| | 2 | DRM Reader (media player, Adobe reader,…) |
| Content server | 3 | Web application |
| | 4 | Database(MS SQL, MySql, Oracle, Postgre,…) |
| | 5 | Smart adapter |
| | 6 | Key generator |
| | 7 | Content encryption |
| License server | 8 | License server |
| | 9 | License generator |
| | 10 | Encryption license |

Concerning the choice of implementations of components either in software or in hardware [28], [29], we held in account the following criteria: the cost in execution time, the type of treatment (management, calculation), the security level, surface (FPGA/SOC/ASIC) or energy consumption. However, we propose the hardware implementation for the component of encoding and authentication of the data based on the IPSec protocol, because it requires an important allocation of the processor because of the complex nature of the calculations made by these operations, in order to offer both a low cost in execution time and a higher level of security.

### 3.3 Components Integration

It is the last phase of the design methodology, which is consists in assembling the components software-software, software-software/hardware or even software-hardware in order to realize a usable system. For the example of the implementation of DRMS, this phase can be summarized in two parts:

#### 3.3.1 Scheduling and parallelism graph

The scheduling and parallelism graph [13], [14] (Fig 5) is a directed graph which represents several possible scenarios of development, so that every scenario represents the same model application and describes the order of execution of components. It includes the connectors of scheduling and parallelisms or one of them to establish indirect connections between components. The main activities ensured by these connectors are [3]: the Sequence, the parallelism, the exclusive choice and the synchronization.

The scheduling and parallelism graph represent all structures *SPG* of the form        *SPG= (FSC, L, δ1, C0, F)* where:
- FSC: Finite set of components.
- $L = SC \cup PC$ where *SC:* scheduling connectors set, PC: parallelism connectors set.
- δ1: EC×L→ f(EC) where f(EC) is the parties set of FSC.
- C0∈ FSC: Initial component.
- F ⊂ FSC: Set of final components.

Several formal techniques can be used to model and validate the GOP: algebras of process, LOTOS or automats, but the technique of modeling which seems particularly suitable to model the GOP is the diagrams of Activity of UML2 [6] because it makes it possible to describe in the form of graph, the sequence of activities and the behavior of the system or its components.

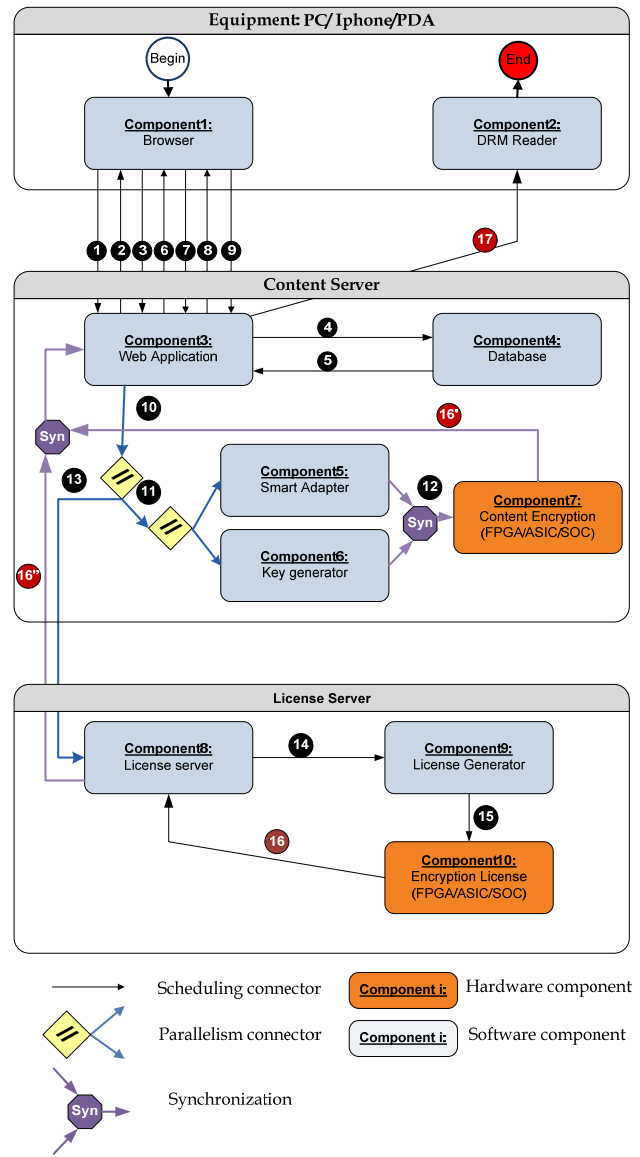

Fig. 5 Schedulling and parallelism graph of DRMS business model

#### 3.3.2 Interaction graph

It is a complementary graph [13], [14] to the scheduling and parallelism graph, its role is to ensure the transfer of data between heterogeneous components (interoperability), this operation is loaded by another type of connector called interaction connector.

The change of the data between the components is made on the level its interfaces, of which if a component (Customer) in an application needs during his execution of a result of another component (supplier), it is necessary to connect the output interface of this last with the input interface of the component (Customer). Of course, it would be necessary to verify that any pair of components to be assembled is compatible of various points of view: syntactic, and interactional.

The interaction graph reprint all structures IG of the form IG= (FSC, IC, δ2) where:
- FSC: Finite set of components.
- IC: set of interaction connectors.
- δ2: FSC × COS × IC 7→f (FSC × CIS).
- COS : Component outputs set.
- CIS : Component inputs set.

## 4. Conclusion

In this article we have presented a unified framework to design software components and hardware components for DRMS systems called F4MS (Framework For Design Mixed Systems). This co-design framework permit to design, develop a high performance, flexible, reuse, maintainable DRM systems with less cost time to deal with rapid growing DRM market.